\RequirePackage{ifpdf}
\ifpdf 
\documentclass[pdftex]{sigma}
\else
\documentclass{sigma}
\fi

\usepackage{euscript}
\usepackage{bbm}
\usepackage{stmaryrd}

\begin{document}
\allowdisplaybreaks

\renewcommand{\PaperNumber}{006}

\FirstPageHeading

\ShortArticleName{Structure of Symmetry Groups via Cartan's Method}

\ArticleName{Structure of Symmetry Groups via Cartan's Method: Survey of
Four Approaches}

\Author{Oleg I. MOROZOV}
\AuthorNameForHeading{O.I. Morozov}

\Address{Moscow State Technical University of Civil Aviation, 125993 Moscow, Russia}
\Email{\href{mailto:oim@foxcub.org}{oim@foxcub.org}}
\URLaddress{\url{http://www.foxcub.org/~oim/}}

\ArticleDates{Received August 08, 2005, in final form September 29, 2005; Published online October 13, 2005}

\Abstract{In this review article we discuss four recent methods for computing
Maurer--Cartan structure equations of symmetry groups of differential equations.
Examples include solution of the contact equivalence problem for linear hyperbolic
equations and finding a~contact transformation between the generalized Hunter--Saxton
equation and the Euler--Poisson equation.}


\Classification{58H05;~58J70;~35A30}

\section{Introduction}

The theory of Lie  groups has greatly influenced diverse branches of
ma\-the\-ma\-tics and physics. The main tool of the theory, Sophus Lie's
infinitesimal method~\cite{Lie}, establishes connection between
con\-ti\-nu\-ous transformation groups and algebras of their infinitesimal
generators. The method leads to many techniques of great significance in
studying the group-invariant solutions and con\-ser\-va\-ti\-on laws of
differential equations~\cite{Ovsiannikov,Ibragimov,Olver86,BlumanKumei,KLV}.
Application of Lie's in\-fi\-ni\-te\-si\-mal method to concrete systems of
differential  equations requires analysis and integration of over-determined
{\it defining systems} for symmetry algebras. Additional integrations need to
be done when differential invariants and operators of invariant differentiations
are computed via the infinitesimal method. Also, complexity of defining
systems in symmetry analysis of clas\-ses of differential equations sometimes
is high enough to make the full study of all branches of classification trees
very hard.

An alternative approach for studying Lie (pseudo-)groups was developed by
{\'E}lie Cartan~\cite{Cartan2,Cartan4,Cartan5}. His theory is based on
characterizing trans\-for\-ma\-ti\-ons from a pseudo-group by means of a set
of invariant differential 1-forms called {\it Maurer--Cartan forms}.
Expressions of exterior differentials of Maurer--Cartan forms in terms of the
forms themselves yield the {\it Cartan structure equations} for the
pseudo-group. These equations contain all information about the pseudo-group,
in particular, they give all differential invariants and operators of invariant
differentiations. The knowledge of Maurer--Cartan forms and differential
invariants enables one to solve equivalence problems for classes of
differential equations  and to find mappings between equivalent equations.
An important feature of the method is that it does not require integration, and
allows one to  find Maurer--Cartan forms by means of only differentiation
and linear algebra operations.

Nowadays there are different methods for computing Maurer--Cartan forms and
struc\-ture  equa\-ti\-ons of symmetry pseudo-groups of differential equations.
In this review paper we discuss four of these methods.
We restrict our attention to symmetries of partial differential equations only.
For applications of Cartan's method to symmetries of ordinary differential
equations we refer, e.g., to
\cite{KamranLambShadwick,KamranShadwick_ODE,Kamran,GrissomThompsonWilkens,%
HsuKamran,BryantGriffithsHsuII,Olver95,Fels,FelsOlver}.

\section{Apriori known geometric properties of a given
differential equation}

The first  approach is presented in \cite{Surovikhin,KamranShadwick,Kamran,%
BryantGriffiths,BryantGriffithsHsuI,BryantGriffithsHsuII,Clelland,Foltinek}.
It is based on apriori known in\-for\-ma\-ti\-on about geometric properties of
a given differential equation. We illustrate the method on the example of
Liouville's equation
\begin{gather}
u_{xy} = e^u.
\label{LiouvilleEquation}
\end{gather}
It is a hyperbolic equation of second order in two independent variables.

From the geometric theory of such equations (see, e.g.,~\cite{BryantGriffithsHsuII}) it follows that
 there exists a coframe (a collection of independent 1-forms)
$\vartheta$, $\omega^i$, $i \in \{1,\ldots,4\}$, on $\mathbb{R}^5$ with
coordinates $(x, y, u, p, q)$, such that a local section
$\sigma : (x, y) \mapsto (x, y, u(x,y), p(x,y), q(x,y))$
of the trivial bundle $\mathbb{R}^5 \rightarrow \mathbb{R}^2$,
$(x, y, u, p, q) \mapsto (x, y)$,
provides a local solution $u(x, y)$ of equation (\ref{LiouvilleEquation})
whenever $\sigma^{*}\,\vartheta=0$,
$\sigma^{*}\,(\omega^1 \wedge \omega^2)=0$,
and $\sigma^{*}\,(\omega^3 \wedge \omega^4)=0$. Indeed, we take
\[
\vartheta  = du - p\, dx - q\, dy,
\quad
\omega^1 = dp - e^u\,dy,
\quad
\omega^2 = dx,
\quad
\omega^3 = dq - e^u\,dx,
\quad
\omega^4 = dy.
\]
Then a local diffeomorphism $\Phi :\mathbb{R}^5 \rightarrow \mathbb{R}^5$,
$\Phi : (x,y,u,p,q) \mapsto (\overline{x},\overline{y},\overline{u},
\overline{p},\overline{q})$,
is a contact symmetry of equation (\ref{LiouvilleEquation}) whenever it
preserves the ideal generated by the contact form $\vartheta$ and
the ideals generated by the pairs $\omega^1$, $\omega^2$ and $\omega^3$,
$\omega^4$ modulo the contact ideal. That is, $\Phi$ must satisfy the
following condition:
\[
\Phi^{*}\,
\left(
\begin{array}{l}
\overline{\vartheta}
\\
\overline{\omega}^1
\\
\overline{\omega}^2
\\
\overline{\omega}^3
\\
\overline{\omega}^4
\end{array}
\right)
=
\left(
\begin{array}{lllll}
a & 0 & 0 & 0 & 0 \\
c_1 & b^1_1 & b^1_2 & 0 & 0 \\
c_2 & b^2_1 & b^2_2 & 0 &  0 \\
c_3 & 0 & 0 & b^3_3 & b^3_4 \\
c_4 & 0 & 0 & b^4_3 & b^4_4 \\
\end{array}
\right)
\,
\left(
\begin{array}{l}
\vartheta
\\
\omega^1
\\
\omega^2
\\
\omega^3
\\
\omega^4
\end{array}
\right)
\]
with $a\big(b^1_1\,b^2_2-b^1_2\,b^2_1\big)\big(b^3_3\,b^4_4-b^3_4\,b^4_3\big)
\not = 0$.

This is a standard set-up for Cartan's equivalence method
\cite{Cartan2,Cartan4,Cartan5,Gardner,Kamran,Olver95}.
Procedures of this method give the Maurer--Cartan forms
\begin{gather*}
\theta^1 = du - p\,dx - q\,dy,
\\
\theta^2 = s_1\,(dp - e^u\,dy) + s_2 \, dx,
\\
\theta^3 = s_1^{-1}\,dx,
\\
\theta^4 = s_1^{-1}\,\left(e^{-u}\,(dq - e^u\,dx)+s_3\,dy\right),
\\
\theta^5  = s_1\,e^u\,dy,
\\
\eta_1 = s_1^{-1}\,d s_1 + p\,dy,
\\
\eta_2 = s_1\,ds_2 - s_2\,ds_1 + s_1^2\, p\, dp
- s_1^{-1}\,(s_1^2\,s_2\,p -z_1)\,dx,
\\
\eta_3 = e^{-2 u}\,s_1^{-2}\,\left(s_1\,ds_3 + s_3\,ds_1+q\,dq
+s_1\,(s_3 +s_1^2\,z_2)\,dy\right),
\end{gather*}
where $s_1=b^1_1$, $s_2 = b^1_2$, $s_3 = b^1_1 b^2_2$, $z_1$
and $z_2$ are arbitrary parameters,  with the structure equations
\begin{gather}
d\theta^1 = -\theta^2 \wedge \theta^3 - \theta^4\wedge \theta^5,
\nonumber\\
d\theta^2 = \eta_1 \wedge \theta^2 +\eta_2 \wedge \omega^3
- \theta^1 \wedge \theta^5,
\nonumber\\
d\theta^3 = -\eta_1 \wedge \theta^3,
\nonumber\\
d\theta^4 = - \eta_1 \wedge \theta^2 +\eta_3 \wedge \theta^5
- \theta^1\wedge(\theta^3+\theta^4),
\nonumber\\
d\theta^5 = (\eta_1+\theta^1) \wedge \theta^5,
\nonumber\\
d\eta_1 = - \theta^3 \wedge \theta^4 + \theta^2 \wedge \theta^3,
\nonumber\\
d\eta_2 = \pi_1 \wedge \theta^3 + 2\,\eta_1 \wedge \eta_2,
\nonumber\\
d\eta_3  = \pi_2 \wedge \theta^5 - 2\,(\theta^1 +\eta_1) \wedge \eta_3.\label{L_SE_1}
\end{gather}

Thus the method is quite simple. It does not require writing out defining
systems for symmetry generators. Also, it gives Maurer--Cartan forms,
dif\-fe\-ren\-ti\-al invariants, and invariant de\-ri\-va\-ti\-ves  for
symmetry groups explicitly. But a lot of preliminary work needs to be done
before the method becomes applicable to a given differential equation.
It is neccessary to find a for\-mu\-la\-ti\-on of the equation in terms of
ideals of exterior forms convenient for setting up the equi\-va\-lence problem
for coframes.

\section{Taylor series expansion of defining equations for
infinitesimal generators of transitive Lie pseudo-groups}

The second method was provided in \cite{LisleReidBoulton, LisleReid}. It
extracts information about Cartan's structure equations from defining systems
for infinitesimal generators. The basic point is that in the case of a
transitive symmetry group the explicit form of solutions of the defining system
is not required for computing coefficients of the structure equations. These
coefficients depend on the finite order Taylor series expansions of the
generators, while the information about the Taylor series can be
obtained from the involutive form of the defining system.

When the defining system for the infinitesimal generator
$X = \sum \limits_{i=1}^{n}\xi^i\,\frac{\partial}{\partial x^i}$ of a
transitive Lie pseudo-group $\EuScript{G}$ is of the first order, the procedure
of the method is as follows. Let $\EuScript{P}_1$ be the set of all
parametric derivatives of the defining system. Denote
\begin{gather}
\frac{\partial \xi^i}{\partial x^j} = \phi^\rho,
\qquad
\rho \in \{1, \ldots , \#(\EuScript{P}_1)\},
\label{ParametricDerivatives}
\end{gather}
for all $\frac{\partial \xi^i}{\partial x^j} \in \EuScript{P}_1$. Then the
involutive form of the defining system is
\begin{gather}
\frac{\partial \xi^i}{\partial x^j} =
\sum \limits_{\rho=1}^{\#(\EuScript{P}_1)} A^i_{j \rho}(x) \,\phi^\rho
+\sum \limits_{k=1}^{n} b^i_{jk} (x) \, \xi^k
\label{PrincipalDerivatives}
\end{gather}
for all principal derivaties. As it is shown in \cite{LisleReid}, in this case
Cartan's structure equations of the pseudo-group $\EuScript{G}$ have the form
\[
d \omega^i =
\sum \limits_{\rho=1}^{\#(\EuScript{P}_1)} \sum \limits_{j=1}^{n}
a^i_{j\rho} \,\pi^\rho \wedge \omega^j
+ \sum \limits_{1\le j < k \le n} c^i_{jk}\, \omega^{j} \wedge \omega^{k},
\qquad
i \in \{1, \ldots , n\},
\]
with $a^i_{j\rho} = A^i_{j\rho}(x_0)$ and
$c^i_{jk} = b^i_{kj}(x_0) - b^i_{jk}(x_0)$,
where $x_0$ is any non-singular point  of the defining system
(\ref{PrincipalDerivatives}).

\begin{example} (See \cite[Example 6]{LisleReidBoulton}.)
Consider the Lie pseudo-group on $\mathbb{R}^2$ whose infinitesimal
generator $X=\xi\,\partial_x+\eta\,\partial_y$ satisfies the defining system
\[
\xi_x = {\textstyle{\frac{1}{y}}}\,\eta,
\qquad
\xi_y = 0,
\qquad
\eta_y = {\textstyle{\frac{1}{y}}}\,\eta.
\]
In this system, the only parametric derivative of the first order is $\eta_x$.
We denote it by $\phi^1$ and add the only parametric equation
(\ref{ParametricDerivatives}):
\[
\eta_x =  \phi^1.
\]
Therefore, in any non-singular point $(x_0,y_0)$ with $y_0 \not =0$ we have
$A^2_{11} =1$, $b^1_{12} = y_0^{-1}$,
$b^1_{11} =b^1_{21}=b^1_{22}= b^2_{ij} =0$. This yields the structure
equations
\[
d\omega^1 = -\frac{1}{y_0}\,\omega^1 \wedge \omega^2,
\qquad
d\omega^2 = \pi^1 \wedge \omega^1.
\]
Any value $y_0 \not =0$ is suitable as an initial data point, so we choose
$y_0 = 1$.
\end{example}

Additional work should be done when the order of the defining system is
greater then $1$  \cite[Section~4.4]{LisleReid}.

\begin{example} (See \cite[Example 8]{LisleReidBoulton}.)
For the symmetry group of Liouville's equation (\ref{LiouvilleEquation})
the coefficients of the infinitesimal generator
$X = \xi\,\partial_x +\eta\,\partial_y+\tau\,\partial_u$ satisfy the defining
system in involutive form
\begin{gather}
\tau_{yy}=-\eta_y,
\quad
\eta_{xy} = 0,
\quad
\xi_x = -\tau_y - \eta,
\quad
\xi_y = 0,
\quad
\xi_u = 0,
\quad
\tau_x =0,
\quad
\tau_u = 0,
\quad
\eta_u = 0.\!\!\!
\label{LR_Liouville_def_sys}
\end{gather}
The algorithm of \cite{LisleReid} gives the following form of the Cartan
structure equations:
\begin{gather}
d \omega^1  = -\omega^1 \wedge \omega^6,
\nonumber\\
d \omega^2 = -\omega^2 \wedge \omega^3
+\omega^2 \wedge \omega^6,
\nonumber\\
d \omega^3 = -\omega^1 \wedge \omega^4
- \omega^2 \wedge \omega^5,
\nonumber\\
d \omega^4 = \pi^1 \wedge \omega^1
+ \omega^4 \wedge \omega^6,
\nonumber\\
d \omega^5 = \pi^2 \wedge \omega^2 - \omega^3 \wedge \omega^5
- \omega^5 \wedge \omega^6,
\nonumber\\
d \omega^6 = - \omega^1 \wedge \omega^4.
\label{L_SE_2}
\end{gather}
This result coincides with the previous one: the substitution of
$\theta^1 = \omega_3$, $\theta^2 = -\omega_2-\omega_4$,
$\theta^3 = \omega_1$, $\theta^4 = -\omega_1-\omega_5$,
$\theta^5 = \omega_2$, $\eta_1 = \omega_6$, $\eta_2 = -\pi_1$,
and $\eta_3 = -\pi_2$
into (\ref{L_SE_1}) gives (\ref{L_SE_2}).
\end{example}

The method has the following properties. It is not universal since it is not
applicable to dif\-fe\-ren\-ti\-al equations with intransitive symmetry
pseudo-groups.  Also, it does not give Maurer--Car\-tan forms and
differential invariants explicitly. An integration should be used to find them
from the structure equations~\cite[Section~7.6]{Flanders}.

\section{Invariantized defining equations for Maurer--Cartan forms}

The third method is developed in
\cite{OlverPohjanpeltoI,OlverPohjanpeltoII,ChehOlverPohjanpelto}. It is based
on use of invariantized defining equations for Maurer--Cartan forms of Lie
pseudo-groups.

Let $\EuScript{D}(M)$ be the pseudo-group of local diffeomorphisms $Z\!=\!\phi(z)$
of a manifold $M$, \mbox{$\dim M\! =\! m,\!$} and $\EuScript{D}^{(\infty)}(M)$ be the
bundle of $\infty$-jets of maps in $\EuScript{D}(M)$. Local coordinates of
the base space~$M$ and $\EuScript{D}^{(\infty)}(M)$ are denoted by $z=(z^i)$
and $(z^i, Z^\alpha_I)$, respectively, where $Z=(Z^\alpha)$ are target
coordinates for $\phi \in \EuScript{D}(M)$, $\phi(z) = Z$, and
$Z^\alpha_I = \partial^I Z^\alpha/\partial z^I =
\partial^{i_1+\cdots +i_m} Z^\alpha/(\partial z^1)^{i_1} \cdots (\partial z^m)^{i_m}$
for $I=(i_1, i_2, \ldots , i_m)$ (our notation for multi-indexes differs
slight\-ly from those of
\cite{OlverPohjanpeltoI,OlverPohjanpeltoII,ChehOlverPohjanpelto}).
Then, as it is shown in \cite{OlverPohjanpeltoI}, the Maurer--Cartan forms for
$\EuScript{D}^{(\infty)}(M)$ are $\sigma^\alpha = Z^\alpha_j\,dz^j$,
$\mu^\alpha = dZ^\alpha-Z^\alpha_j\,dz^j$, and
$\mu^\alpha_I = \mathbb{D}^I_Z(\mu^\alpha)$, $\alpha \in \{1,\ldots,m\}$,
$\# I = i_1 + \cdots + i_m \ge 1$,
where
$\mathbb{D}^I_Z
=\mathbb{D}^{i_1}_{Z^1} \circ \cdots \circ\mathbb{D}^{i_m}_{Z^m}$,
\[
\mathbb{D}_{Z^j} = \sum \limits_{i=1}^{m} w^i_j\, \mathbb{D}_{z^i},
\qquad
\mathbb{D}_{z^i} = \frac{\partial}{\partial z^i}+
\sum \limits_{\alpha=1}^{m} \sum \limits_{\# I \ge 0} Z^\alpha_{I+1_i}\,
\frac{\partial}{\partial Z^\alpha_I},
\]
while $(w^i_j)$ is the inverse matrix for the Jacobian matrix $(Z^\alpha_j)$.
The structure equations for $\EuScript{D}^{(\infty)}(M)$ have the form
\begin{gather}
d \mu \llbracket H \rrbracket = \nabla \mu \llbracket H \rrbracket \wedge
\left(\mu \llbracket H \rrbracket - dZ\right),
\label{diffeo_SE_1}
\\
d\sigma = - d \mu \llbracket 0 \rrbracket,
\label{diffeo_SE_2}
\end{gather}
where
$
\mu \llbracket H \rrbracket = \left(\mu^\alpha \llbracket H \rrbracket \right)
= \left(\sum \limits_{\# I \ge 0} \frac{1}{I !} \,\mu^\alpha_I\, H^I\right)
$,
$H^I = H_1^{i_1} \cdots H_m^{i_m}$,
and
$\nabla \mu \llbracket H \rrbracket =
\left(\frac{\partial \mu^\alpha \llbracket H \rrbracket}{\partial H_j}\right)$
denote the Jacobian matrix of the vector $\mu \llbracket H \rrbracket$  of
power series in the variables $H=(H_j)$.

Let $V = \zeta^\alpha \,\partial_{z^\alpha}$ be the infinitesimal generators
of a sub-pseudo-group $\EuScript{G} \subset \EuScript{D}(M)$. They are
cha\-ra\-c\-te\-ri\-zed by the defining equations
\begin{gather}
L(z^i, \zeta^\alpha_I) = 0.
\label{inf_def_sys}
\end{gather}
If $\EuScript{G}$ is the symmetry group of a system of differential equations,
then (\ref{inf_def_sys}) are (the involutive completion of) the usual
determining equations obtained through Lie's infinitesimal technique.

The method is based on the following  theorems \cite{OlverPohjanpeltoI}:

\begin{theorem}
The invariant forms $\mu^\alpha_I$  of a Lie pseudo-group
$\EuScript{G} \subset \EuScript{D}(M)$ satisfy the linear system
\begin{gather}
L(Z^i, \mu^\alpha_I) = 0,
\label{lifted_determining_equations}
\end{gather}
obtained by replacing $z^i$ by $Z^i$ and $\zeta^\alpha_I$ by
$\mu^\alpha_I$ in the determining equations
\eqref{inf_def_sys}.
\end{theorem}

\begin{theorem}
The structure equations of the invariant coframe for a Lie pseudo-group
$\EuScript{G}$ are obtained by restricting the diffeomorphism structure
equations \eqref{diffeo_SE_1}, \eqref{diffeo_SE_2} to the
space of solutions of the equations \eqref{lifted_determining_equations}.
\end{theorem}

These results allow to find the structure equations of the symmetry group of
a system of differential equations from an involutive form of infinitesimal
defining equations. The required computations rely exclusively on linear
algebra and differentiation, and can be readily implemented in any standard
symbolic computation package.

\begin{example}
To illustrate the method, we apply it to Liouville's equation
(\ref{LiouvilleEquation}). We denote $(z^1,z^2,z^3) = (x,y,u)$,
$(Z^1,Z^2,Z^3) = (X,Y,U)$, and take the infinitesimal defining system in the
involutive form (\ref{LR_Liouville_def_sys}). Then from
the fourth and fifth equations of (\ref{LR_Liouville_def_sys})
and their derivatives w.r.t. $y$ and $u$ we have
$\mu^1_{(0,k,l)} = 0$ for $k+l \ge 1$, so
\[
\mu^1 \llbracket H \rrbracket = \sum \limits_{j \ge 0}\,\frac{1}{j!}\,
\mu^1_{(j,0,0)}\,H_1^j.
\]
Further, from the sixth and seventh equations of
(\ref{LR_Liouville_def_sys}) and their derivatives w.r.t.~$x$ and $u$
we have $\mu^2_{(j,0,l)} = 0$ for $j+l \ge 1$, thus
\[
\mu^2 \llbracket H \rrbracket = \sum \limits_{k \ge 0}\,\frac{1}{k!}\,
\mu^2_{(0,k,0)}\,H_2^k.
\]
Finally, from the third equation of (\ref{LR_Liouville_def_sys})
and its derivatives w.r.t.~$x$, $y$, and $u$
we have $\mu^3_{(0,0,0)} = - \mu^1_{(1,0,0)} - \mu^2_{(0,1,0)}$ and
\[
\mu^3 \llbracket H \rrbracket =
- \sum \limits_{j \ge 0}\,\frac{1}{j!}\, \mu^1_{(j+1,0,0)}\,H_1^j
- \sum \limits_{k \ge 0}\,\frac{1}{k!}\, \mu^2_{(0,k+1,0)}\,H_2^k.
\]
Let us denote $\mu^1_{(j,0,0)} = \phi_j$ and $\mu^2_{(0,k,0)} = \psi_k$
for brevity. Then substitution of $\mu^1 \llbracket H \rrbracket$,
$\mu^2 \llbracket H \rrbracket$, and $\mu^3 \llbracket H \rrbracket$ into
(\ref{diffeo_SE_1}) and (\ref{diffeo_SE_2}) yields an infinite system of
structure equations:
\begin{gather*}
\left(
\begin{array}{c}
\sum \limits_{j \ge 0} \frac{1}{j!}\, d \phi_j \, H^j_1
\\
\sum \limits_{k \ge 0} \frac{1}{k!}\, d \psi_k \, H^k_2
\\
-\sum \limits_{j \ge 0} \frac{1}{j!}\, d \phi_{j+1} \, H^j_1
-\sum \limits_{k \ge 0} \frac{1}{k!}\, d \psi_{k+1} \, H^k_2
\end{array}
\right)
\\
= \left(
\begin{array}{ccc}
\sum \limits_{j \ge 0} \frac{1}{j!}\, \phi_{j+1} \, H^j_1 & 0 & 0\\
0 &\sum \limits_{k \ge 0} \frac{1}{k!}\, \psi_{k+1} \, H^k_2 & 0 \\
-\sum \limits_{j \ge 0} \frac{1}{j!}\, \phi_{j+2} \, H^j_1 &
\sum \limits_{k \ge 0} \frac{1}{k!}\, \psi_{k+2} \, H^k_2 & 0
\end{array}
\right)
\wedge
\left(
\begin{array}{ccc}
-\sigma^1+\sum\limits_{j\ge 1} \frac{1}{j!}\, \phi_j\, H^j_1
\\
-\sigma^2+\sum\limits_{k\ge 1} \frac{1}{k!}\, \psi_k\, H^k_2
\\
-\sigma^3-\sum\limits_{j\ge 1} \frac{1}{j!}\, \phi_{j+1}\, H^j_1
-\sum\limits_{k\ge 1} \frac{1}{k!}\, \psi_{k+1}\, H^k_2
\end{array}
\right)\!\!\!
\end{gather*}
and
\[
d\sigma^1 = - d\phi_0,
\qquad
d\sigma^2 = - d\psi_0,
\qquad
d\sigma^3 = d\phi_1 + d\psi_1.
\]

From these equations we have
\begin{gather}
d\phi_0 = -\phi_1 \wedge \sigma^1,
\nonumber\\
d\phi_j = - \phi_{j+1} \wedge \sigma^1
+ \sum \limits_{p\ge 0, q\ge 1, p+q = j}
\frac{j!}{p! q!} \, \phi_{p+1}\wedge \phi_q,
\qquad j \ge 1,
\nonumber\\
d\psi_0 = -\psi_1 \wedge \sigma^2,
\nonumber\\
d\psi_k = - \phi_{k+1} \wedge \sigma^2
+ \sum \limits_{p\ge 0, q\ge 1, p+q = k}
\frac{k!}{p! q!} \, \psi_{p+1}\wedge \psi_q,
\qquad k \ge 1,
\nonumber\\
d\sigma^1 = \phi_1 \wedge \sigma^1,
\nonumber\\
d\sigma^2 = \psi_1 \wedge \sigma^2,
\nonumber\\
d\sigma^3 = -\phi_2 \wedge \sigma^1 - \psi_2 \wedge \sigma^2.
\label{L_SE_3}
\end{gather}

To establish correspondence with the previous results, we note that the
substitution of
$\sigma^1=\omega^1$,
$\sigma^2=\omega^2$,
$\phi_1 =\omega^6$,
$\phi_2 = \omega^4$,
$\psi_1 = \omega^3-\omega^6$,
$\psi_2 = \omega^5$,
$\phi_3 = -\pi_1$,
and
$\psi_3 = -\pi_2$
into equations (\ref{L_SE_3}) for $d\sigma_1$,  $d\sigma_2$, $d\phi_1$,
$d\phi_2$, $d\psi_1$, and $d\psi_2$ gives equations~(\ref{L_SE_2}),
while equations~(\ref{L_SE_3}) for $d\phi_{i+1}$ and $d\psi_{k+1}$
appear from the exterior differentials of the equations for $d\phi_{i}$ and
$d\psi_{k}$, respectively. Therefore, equations (\ref{L_SE_3}) are 
infinite prolongation of equations (\ref{L_SE_2}).

The sets of equations (\ref{L_SE_3}) for $\sigma^1$, $\phi_i$, and
$\sigma^2$, $\psi_k$ are independent. Moreover, each of these sets coincides
with the structure equations for the infinite prolongation of the
diffeomorphism pseudo-group on $\mathbb{R}^1$~\cite{Cartan1},
\cite[Example~4.1]{OlverPohjanpeltoI}. This shows that the pseudo-group of contact
symmetries of Liouville's equation is a direct product of two diffeomorphism
pseudo-groups on $\mathbb{R}^1$, as it follows, of course, from the results
obtained by the infinitesimal method \cite[Bd.~5, 469--478]{Lie}.
\end{example}

Unlike the previous two methods, the method of
\cite{OlverPohjanpeltoI,OlverPohjanpeltoII,ChehOlverPohjanpelto} is universal
since it is applicable to any differential equation. It requires analysis of
the defining systems for infinitesimal generators and its reduction to the
involutive form. For differential equations with infinite symmetry
pseudo-groups the method produces infinite sets of Maurer--Cartan forms and
infinite systems of structure equations. Also, the Maurer--Cartan forms
obtained by this method depend on both source and target variables of the
diffeomorphism pseudo-group. Therefore, additional work needs to be done to
obtain finite sets of Maurer--Cartan forms and to express them in terms which
are suitable for further implementations such as finding transformations
between equivalent equations.

\section{The moving coframe method}

Finally, the fourth approach is based on the moving coframe method provided in
\cite{FelsOlver}. Applied to contact symmetries of differential equations of the
second order, the method has the following outline. Consider the bundle
$J^2(\EuScript{E})$ of the second order jets of local sections of the trivial
bundle
$\EuScript{E} = \mathbb{R}^n \times \mathbb{R} \rightarrow \mathbb{R}^n$.
Contact transformations
$\Delta : J^2(\EuScript{E}) \rightarrow J^2(\EuScript{E})$,
$\Delta : (x^i,u,p_i,p_{ij}) \mapsto
(\overline{x}^i,\overline{u},\overline{p}_i,\overline{p}_{ij})$,
are characterized by requirements
$\Delta^{*}(d\overline{u}-\overline{p}_i\,d\overline{x}^i)
\equiv 0 \,\,\,({\mathrm{mod}}\,\,\, du-p_i\,dx^i)$ and
$\Delta^{*}(d\overline{p}_i-\overline{p}_{ij}\,d\overline{x}^j)
\equiv 0 \,\,\,({\mathrm{mod}}\,\,\, du-p_i\,dx^i,\, dp_i-p_{ij}\,dx^j)$.
Therefore,  Maurer--Cartan forms for the pseudo-group $\mathop{\rm Cont}(J^2(\EuScript{E}))$
of contact transformations on $J^2(\EuScript{E})$ can be easily found using
Cartan's equivalence method, see, e.g., \cite{Morozov2004}. These forms are
\begin{gather*}
\Theta_0 = a\,(du-p_i\,dx^i),
\\
\Theta_i = g_i\,\Theta_0+ a\,B^k_i\,(dp_k-p_{kl}\,dx^l),
\\
\Xi^i = c^i\,\Theta_0+f^{ik}\,\Theta_k+b^i_k\,dx^k,
\\
\Sigma_{ij} = s_{ij}\,\Theta_0+w^{k}_{ij}\,\Theta_k+z_{ijk}\,\Xi^k
+a\,B^k_i\,B^l_j\,dp_{kl},
\end{gather*}
where $i, j \in \{1,\ldots,n\}$, $a\not = 0$, $\mathrm{det}\,(b^i_j) \not = 0$,
$f^{ik}=f^{ki}$, $s_{ij} = s_{ji}$, $w^k_{ij} = w^k_{ji}$,
$z_{ijk} = z_{jik} = z_{ikj}$, while $(B^i_j)$ is the inverse matrix for the
matrix $(b^i_j)$. The structure equations of $\mathop{\rm Cont}\nolimits (J^2(\EuScript{E}))$ have the
following form:
\begin{gather*}
d \Theta_0 = \Phi^0_0 \wedge \Theta_0 + \Xi^i \wedge \Theta_i,
\\
d \Theta_i = \Phi^0_i \wedge \Theta_0 + \Phi^k_i \wedge \Theta_k
+ \Xi^k \wedge \Sigma_{ik},
\\
d \Xi^i = \Phi^0_0 \wedge \Xi^i -\Phi^i_k \wedge \Xi^k
+\Psi^{i0} \wedge \Theta_0
+\Psi^{ik} \wedge \Theta_k,
\\
d \Sigma_{ij} = \Phi^k_i \wedge \Sigma_{kj} - \Phi^0_0 \wedge \Sigma_{ij}
+ \Upsilon^0_{ij} \wedge \Theta_0
+ \Upsilon^k_{ij} \wedge \Theta_k + \Lambda_{ijk} \wedge \Xi^k.
\end{gather*}

A differential equation $\EuScript{R}$ of the second order is a subbundle of
$J^2(\EuScript{E})$. Let  $\iota : \EuScript{R} \rightarrow J^2(\EuScript{E})$
be the inclusion map. Then we can find the Maurer--Cartan forms for the
pseudo-group $\mathop{\rm Sym}\nolimits (\EuScript{R})$ of contact symmetries of $\EuScript{R}$
from the restrictions
$\theta_0 = \iota^{*}\,\Theta_0$,
$\theta_i = \iota^{*}\,\Theta_i$,
$\xi^i = \iota^{*}\,\Xi^i$,
$\sigma_{ij} = \iota^{*}\,\Sigma_{ij}$
by standard procedures of Cartan's equivalence method, see
\cite{FelsOlver,Morozov2002,Morozov2003} for details.

\begin{example}
Applying the moving coframe method to the symmetry pseudo-group of Liouville's
equation  (\ref{LiouvilleEquation}), we obtain the Maurer--Cartan forms
\begin{gather*}
\theta_0=du-u_x dx - u_y dy,
\\
\theta_1=r_1^{-1}\,(du_x - u_{xx} dx - e^u dy),
\\
\theta_2=r_1\,e^{-u}\,(du_y - e^u dx-u_{yy} dy),
\\
\xi^1 = r_1 dx,
\\
\xi^2 = r_1^{-1} e^u dy,
\\
\sigma_{11} =r_1^{-2}\,\left(du_{xx} - u_x du_x
+(u_x u_{xx} + r_1^3 r_2)\,dx\right),
\\
\sigma_{22} = r_1^2 e^{-2u}\,\left(du_{yy}- u_y du_y
+(u_y u_{yy} + r_1^{-3} r_3)dy\right),
\\
\eta_1=r_1^{-1}(dr_1 - u_x\,\xi^1),
\\
\eta_2=dr_2 - 3\,r_2\,\eta_1+r_1^{-2}(u_{xx}+u_x^2)\,(\theta_1+\xi^2)
+3\,r_1^{-1}u_x\,\sigma_{11}+v_1\,\xi^1,
\\
\eta_3=dr_3+3\,r_3\,(\eta_1+\theta_0)
+r_1^2e^{-2u}\,(u_{yy}+u_y^2)\,(\theta_2+\xi^1)
+3\,r_1e^{-u} u_y \sigma_{22}+v_2\,\xi^2,
\end{gather*}
where $r_1 = b^1_1$, $r_2 = z_{111}$, $r_3 = z_{222}$, $v_1$ and $v_2$
are arbitrary parameters, while $\sigma_{12} = 0$.
The forms satisfy the following structure equations:
\begin{gather}
d\theta_0 = -\theta_1 \wedge \xi^1 - \theta_2 \wedge \xi^2,
\nonumber\\
d\theta_1 = \eta_1 \wedge \theta_1 - \theta_0 \wedge \xi^2
+\xi^1 \wedge \sigma_{11},
\nonumber\\
d\theta_2 = -\eta_1 \wedge \theta_2 - \theta_0 \wedge (\theta_2+\xi^1)
+\xi^2 \wedge \sigma_{22},
\nonumber\\
d\xi^1 = -\eta_1 \wedge \xi^1,
\nonumber\\
d\xi^2 = (\eta_1+\theta_0) \wedge \xi^2,
\nonumber\\
d\sigma_{11} = \eta_2 \wedge \xi^1+2\,\eta_1 \wedge \sigma_{11},
\nonumber\\
d\sigma_{22} = \eta_3 \wedge \xi^2-2\,(\eta_1+\theta_0) \wedge \sigma_{22},
\nonumber\\
d\eta_1  = (\theta_1+\xi^2) \wedge \xi^1,
\nonumber\\
d\eta_2 = \pi_1 \wedge \xi^1+3\,\eta_1 \wedge \eta_2
+2\,(\theta_1+\xi^2) \wedge \sigma_{11},
\nonumber\\
d\eta_3 = \pi_2 \wedge \xi^2-3\,(\eta_1+\theta_0) \wedge \eta_3
+2\,(\theta_2+\xi^1) \wedge \sigma_{22}.
\label{L_SE_4}
\end{gather}

The substitution of $\xi^1 = \omega^1$, $\xi^2 = \omega^2$,
$\theta_0 = \omega^3$, $\theta_1 = -\omega^2-\omega^4$,
$\theta_2 = -\omega^1-\omega^5$, $\eta_1 = -\omega^6$,
$\eta_2 = \pi_1$, and $\eta_3 = \pi_2$ into (\ref{L_SE_4}) yields equations
(\ref{L_SE_2}). Therefore, this result coincides with the results obtained in
previous examples.
\end{example}

The moving coframe method enables one to find Maurer--Cartan forms,
differential invariants, and operators of invariant differentiations for
symmetry pseudo-groups explicitly, in contrast to the method of
\cite{LisleReid}. It does not require integration at all, and, unlike the
methods of
\cite{LisleReid,OlverPohjanpeltoI,OlverPohjanpeltoII,ChehOlverPohjanpelto},
does not use infinitesimal defining systems. Also, the method is universal
since it is applicable to any differential equation. The price of these
advantages is lengthy and intricate computations. Ne\-ver\-the\-less the
method allows one to solve effectively problems concerned with symmetry
clas\-si\-fi\-ca\-ti\-on and equivalence of differential equation. We illustrate
it in the two following sections.

\section{Contact equivalence problem for linear hyperbolic equations}

In \cite{Morozov2004} the moving coframe method is used to solve the local
equivalence problem for the class of linear second order hyperbolic equations
in two independent variables
\begin{gather}
u_{tx} = T(t,x)\,u_t+X(t,x)\,u_x+U(t,x)\,u
\label{LHE}
\end{gather}
w.r.t.\ the pseudo-group of contact transformations.

This class has been studied for more than two centuries. In \cite{Laplace}
P.S.~Laplace found semi-invariants $H = - T_t+T\,X+U$ and $K=-X_x+T\,X+U$.
These functions are  invariants of the sub-pseudo-group
$u\mapsto \lambda(t,x) u$,  $\lambda \not = 0$. Laplace proved that equation
(\ref{LHE}) is equivalent to the wave equation $u_{tx} = 0$ w.r.t.\ this
sub-pseudo-group whenever $H\equiv 0$ and $K \equiv 0$.

S.~Lie studied equations (\ref{LHE}) by means of the infinitesimal method
\cite[Bd.~3,~492--523]{Lie}. He found ca\-no\-ni\-cal forms for equations
(\ref{LHE}) and methods of their integration.

L.V.~Ovsiannikov found contact invariants $P=K/H$,
$Q=(\ln \vert \,H\,\vert\,)_{tx}/H$,  and  applied them to the problem of
classification of equations (\ref{LHE}) with non-trivial finite-dimensional
sub\-gro\-ups of symmetry pseudo-groups \cite{Ovsiannikov1960},
\cite[Section~9.2]{Ovsiannikov}.

The solution of the contact equivalence problem for the class
(\ref{LHE}) is found in
\cite{Morozov2004}:
\begin{theorem}
Class \eqref{LHE} is divided into the six sub\-clas\-ses
$\EuScript{C}_1, \EuScript{C}_2, \ldots, \EuScript{C}_6$ invariant under
an action of the pseudo-group of contact transformations:
\begin{enumerate}
\itemsep=0pt
\item[$\EuScript{C}_1$] consists of all equations \eqref{LHE}   such that
$H \equiv 0$ and $K \equiv 0$;
\item[$\EuScript{C}_2$] consists of all equations \eqref{LHE} such that
$P_t \not= 0$;
\item[$\EuScript{C}_3$] consists of all equations \eqref{LHE} such that
$P_t \equiv 0$ and
$P_x \not = 0$;
\item[$\EuScript{C}_4$] consists of all equations \eqref{LHE} such that
$P \equiv {\rm const}$ and $Q_t \not = 0$;
\item[$\EuScript{C}_5$]  consists of all equations \eqref{LHE} such that
$P \equiv {\rm cons}t$, $Q_t \equiv 0$, and $Q_x \not = 0$;
\item[$\EuScript{C}_6$]  consists of all equations \eqref{LHE} such that
$P \equiv{\rm  const}$ and $Q \equiv {\rm const}$.
\end{enumerate}

Every equation from the subclass $\EuScript{C}_1$  is equivalent to the linear
wave equation $u_{tx} = 0$.

Every equation from the subclass $\EuScript{C}_6$ is equivalent to the
equation
\[
u_{tx} = -t\,u_t-P\,x\,u_x-P\,t\,x\,u
\]
when $Q = 0$,  or to the Euler--Poisson equation
\begin{gather}
u_{tx} = \frac{2}{Q\,(t+x)}\, u_t + \frac{2\,P}{Q\,(t+x)}\, u_x
- \frac{4\,P}{Q^2\,(t+x)^2}\,u
\label{EulerPoisson_general}
\end{gather}
when $Q\not = 0$.

For the subclass  $\EuScript{C}_2$, the basic invariants are  $P$, $Q$,  and
$J = (H_tP_t - H P_{tt})\,H^{-1} P_t^{-2}$,
the operators of  invariant differentiation are
$\mathbb{D}_1 = P_t^{-1} D_t$ and
$\mathbb{D}_2 = P_t H^{-1} D_x$.

For $\EuScript{C}_3$, the basic invariants are
$P$, $Q$, and
$L = (H_xP_x - H P_{xx})\,H^{-1} P_x^{-2}$,
the operators of  invariant differentiation are
$\mathbb{D}_1 = P_x H^{-1} D_t$ and
$\mathbb{D}_2 = P_x^{-1} D_x$.

For $\EuScript{C}_4$,  the basic invariants are
$Q$, $M_1 = Q_{tx}H^{-1}$, and
$M_2 = (H_tQ_t - H Q_{tt})\,H^{-1} Q_t^{-2}$,
the operators of  invariant differentiation are
$\mathbb{D}_1 = Q_t^{-1} D_t$ and
$\mathbb{D}_2 = Q_t H^{-1} D_x$.

For 
$\EuScript{C}_5$, the basic invariants are
$Q$ and
$N = (H_xQ_x - H Q_{xx})\,H^{-1} Q_x^{-2}$,
the operators of  invariant differentiation are
$\mathbb{D}_1 = Q_x H^{-1} D_t$ and
$\mathbb{D}_2 = Q_x^{-1} D_x$.

Two equations from the subclasses $\EuScript{C}_2$--$\EuScript{C}_5$
are locally equivalent to each other if and only if they have the same
functional dependencies among the basic invariants and their invariant
derivatives up to the second order.
\end{theorem}

Different results were stated in \cite{JohnpillaiMahomedWafoSoh} and
\cite{Ibragimov2004}:

\begin{theorem}[\cite{JohnpillaiMahomedWafoSoh}]
\label{JohnpillaiMahomedWafoSoh_th}
The functions $P$, $Q$,
\[
J^1_3 = H^{-3}\,\left(K\,H_{tx}+H\,K_{tx}-H_tK_x-H_xK_t\right),
\]
\[
J^2_3 = H^{-9}\,\left(H\,K_x-K\,H_x\right)^2\,
\left(H\,K\,H_{tt}-H^2K_{tt}-3\,K\,H^2_t+3\,H\,H_tK_t\right),
\]
and
\[
J^3_3 = H^{-9}\,\left(H\,K_t-K\,H_t\right)^2\,
\left(H\,K\,H_{xx}-H^2K_{xx}-3\,K\,H^2_x+3\,H\,H_xK_x\right)
\]
are a basis of the complete set of invariants of \eqref{LHE}. Any other
differential invariant is a function of these basic invariants and their
invariant derivatives.
\end{theorem}

\begin{theorem}[\cite{Ibragimov2004}]
\label{Ibragimov2004_th}
A basis of invariants for equations \eqref{LHE} consists of the invariants
\[
P, \quad
Q, \quad
I = P_tP_xH^{-1}, \quad
\widetilde{Q} = (\ln \vert\,K\,\vert\,)_{tx}K^{-1},
\]
or of invariants of the alternative basis
\[
P, \quad
Q, \quad
I, \quad
J = (H_tP_t - H P_{tt})\,H^{-1} P_t^{-2}.
\]
\end{theorem}

The operators of invariant differentiations are taken in \cite{Ibragimov2004}
in the following form: $\EuScript{D}_1= P_t^{-1}D_t$ and
$\EuScript{D}_2 = P_x^{-1}D_x$.

The functions of Theorems \ref{JohnpillaiMahomedWafoSoh_th},
\ref{Ibragimov2004_th} do not provide bases of invariants for the {\it whole}
class of equations (\ref{LHE}). To prove this, we consider the following
equation:
\begin{gather}
u_{tx} = u_t + \frac{2\,(p(t)-1)}{q(t)\,(t+x)}\,u_x
+ \frac{2\,\left(1-(p(t)-1)\,(t+x)\right)}{q(t)\,(t+x)^2}\,u
\label{counter_example}
\end{gather}
with
$p(t) \not = 0$, $q(t) \not = 0$, $p^{\prime}(t) \not = 0$, and
$q^{\prime}(t) \not = 0$.
For this equation we have
$P=p(t)$,
$Q=q(t)$,
$J^1_3 = 2\,p(t)\,q(t)$,
$J^2_3 = J^3_3 = I =0$,
$\widetilde{Q} = q(t)\,(p(t))^{-1}$,
while
\[
J = -\frac{2}{q^{\prime}(t)\,(t+x)}
- \frac{p^{\prime \prime}(t)}{(p^{\prime}(t))^2}
- \frac{q^{\prime}(t)}{p^{\prime}(t)\,q(t)}.
\]
Since $J$ depends on $x$ explicitly, it is functionally independent of $P$,
$Q$, $J^1_3$, $J^2_3$, $J^3_3$, $I$, and $\widetilde{Q}$.  Therefore, the
functions from Theorem~\ref{JohnpillaiMahomedWafoSoh_th} and the first set of
functions from Theorem~\ref{Ibragimov2004_th} do not provide a basis of
differential invariants for the whole class (\ref{LHE}).

Otherwise, for Moutard's equation
\begin{gather}
u_{tx} = U(t,x)\, u
\label{Moutard}
\end{gather}
we have  $\widetilde{Q} = Q$, $P=1$, so $P_t = P_x = I= 0$. Therefore the
invariant $J$ and the operators $\EuScript{D}_1= P_t^{-1}D_t$,
$\EuScript{D}_2 = P_x^{-1}D_x$ are not defined for this equation. This proves
that the both sets of functions from Theorem~\ref{Ibragimov2004_th} do not
provide bases of invariants for the subclass (\ref{Moutard}).

\begin{remark}
As it is shown in \cite{Morozov2004}, the symmetry pseudo-groups of equations
(\ref{LHE}) from the sub\-clas\-ses $\EuScript{C}_2$--$\EuScript{C}_5$ are
intransitive. Therefore, the moving coframe method is shown to be
ap\-pli\-ca\-ble to study intransitive symmetry pseudo-groups as well.
\end{remark}

\section{Linearizability of the generalized Hunter--Saxton equation}

In this section, we use the moving coframe method to study the generalized
Hunter--Saxton equa\-ti\-on
\begin{gather}
u_{tx} = u\,u_{xx}+\kappa\,u_x^2.
\label{HS}
\end{gather}
This equation has a number of applications in the nonlinear instability theory
of a director field of a liquid crystal~\cite{HunterSaxton}, in geometry of
Einstein--Weil spaces~\cite{Tod,Dryuma}, in constructing partially invariant
solutions for the Euler equations of an ideal fluid~\cite{Golovin}, and has
been a subject of many recent investigations. In the case
$\kappa = \frac{1}{2}$
the general solution~\cite{HunterSaxton}, the tri-Hamiltonian formulation~\cite{OlverRosenau},
the pseudo-spherical formulation and the quadratic
pseudo-potentials~\cite{Reyes} have been found. The conjecture of
linearizability of equa\-ti\-on (\ref{HS}) in the case $\kappa = - 1$ has been
made in \cite{Golovin}. In \cite{Pavlov} a formula for the general solution of
(\ref{HS}) has been proposed. This formula uses a nonlocal change of variables.

We prove that equation (\ref{HS}) is equivalent under a contact
trans\-for\-ma\-ti\-on to the Euler--Poisson equation
(\ref{EulerPoisson_general}) with $P= 2\,(1-\kappa)$ and $Q = 2\,\kappa$:
\begin{gather}
u_{tx} = \frac{1}{\kappa\,(t+x)}\,u_t
+ \frac{2\,(1-\kappa)}{\kappa\,(t+x)}\,u_x
- \frac{2\,(1-\kappa)}{(\kappa\,(t+x))^{2}}\,u.
\label{EP}
\end{gather}
Also, we find the general solution of (\ref{HS}) in terms of local variables.

Using the moving coframe method, we obtain the structure equa\-ti\-ons for the
symmetry pseudo-group of equa\-ti\-on (\ref{EP}) in the form
\begin{gather}
d\theta_0 =
\eta_1\wedge\theta_0
+\xi^1\wedge\theta_1
+\xi^2\wedge\theta_2,
\nonumber\\
d\theta_1 =
\eta_2\wedge\theta_1
-2\,(1-\kappa)\,\theta_0\wedge\xi^2
+\xi^1\wedge\sigma_{11},
\nonumber\\
d\theta_2 =
(2\,\eta_1-\eta_2)\wedge\theta_2
-\theta_0\wedge\xi^1
+\xi^2\wedge\sigma_{22},
\nonumber\\
d\xi^1 =
(\eta_1-\eta_2)\wedge\xi^1,
\nonumber\\
d\xi^2 =
(\eta_2 -\eta_1)\wedge\xi^2,
\nonumber\\
d\sigma_{11} =
(2\,\eta_2-\eta_1)\wedge\sigma_{11}
+\eta_3\wedge\xi^1
+3\,(2\,\kappa-1)\,\theta_1\wedge\xi^2,
\nonumber\\
d\sigma_{22} =
(3\,\eta_1-2\,\eta_2)\wedge\sigma_{22}
+\eta_4\wedge\xi^2,
\nonumber\\
d\eta_1 =
(2\,\kappa-1)\,\xi^1\wedge\xi^2,
\nonumber\\
d\eta_2 =
(1-4\,\kappa)\,\xi^1\wedge\xi^2,
\nonumber\\
d\eta_3 =
\pi_1\wedge\xi^1
- (2\,\eta_1-3\,\eta_2)\wedge\eta_3
+4\,(3\,\kappa-1)\,\xi^2\wedge\sigma_{11},
\nonumber\\
d\eta_4 =
\pi_2\wedge\xi^2
+(4\,\eta_1-3\,\eta_2)\wedge\eta_4
+2\,(3-\kappa)\,\xi^1\wedge\sigma_{22},
\label{SE}
\end{gather}
with
$\theta_0 = a\,(d u-u_t\,d t-u_x\,d x)$,
$\theta_1 = a\,b^{-1}(d u_t - u_{tt}d t -R_2\,d x)
+ 2\,(\kappa-1)\,(\kappa\,b\,(t+x))^{-1}\theta_0$,
$\theta_2 =a\,b\,\kappa\,(t+x)^2\,(d u_x - R_2\,d t- u_{xx}\,d x)
+ b\,(t+x)\,\theta_0$,
$\xi^1 = b\,d t$, and $\xi^2 = b^{-1} \kappa^{-1} (t+x)^{-2} d x$,
where $R_2$ is the right-hand side of equation (\ref{EP}),
while $a$ and  $b$ are arbitrary non-zero constants.
The forms $\sigma_{11}, \ldots , \pi_2$ are too long to be written out
in full here.
We write equation (\ref{HS}) and its Maurer--Cartan forms in tilded
variables, then similar computations give
$\widetilde{\theta}_0 = \widetilde{a}\,(d\widetilde{u}
-\widetilde{u}_{\widetilde{t}}\,d\widetilde{t}
-\widetilde{u}_{\widetilde{x}}\,d\widetilde{x})$,
$\widetilde{\theta}_1 = \widetilde{a}\,\widetilde{b}^{-1}
(d\widetilde{u}_{\widetilde{t}}
- \widetilde{u}_{\widetilde{t}\widetilde{t}}\,d\widetilde{t}
- {\widetilde{R}}_1\,d\widetilde{x})
- \widetilde{b}^{-2} \widetilde{u}\,
\widetilde{u}_{\widetilde{x}\widetilde{x}}\,\, \widetilde{\theta}_2
- (2\,\kappa-1)\,\widetilde{b}\,\widetilde{u}_{\widetilde{x}}\,\,
\widetilde{\theta}_0$,
$\widetilde{\theta}_2 = \widetilde{a}\,
\widetilde{b}^{-1}(\widetilde{u}_{\widetilde{x}\widetilde{x}})^{-1}\,
(d\widetilde{u}_{\widetilde{x}}-{\widetilde{R}}_1
\,d\widetilde{t}
- \widetilde{u}_{\widetilde{x}\widetilde{x}}\,d\widetilde{x})$,
$\widetilde{\xi}^1 = \widetilde{b}\,d\widetilde{t}$,
and
$\widetilde{\xi}^2 = \widetilde{b}^{-1}\,(d\widetilde{u}_{\widetilde{x}}
- \kappa\,(\widetilde{u}_{\widetilde{x}})^2\,d\widetilde{t})$,
where ${\widetilde{R}}_1$ is the right-hand side of equation (\ref{HS})
written in the tilded vatiables, while ${\widetilde{a}}$ and  ${\widetilde{b}}$
are arbitrary non-zero constants. The forms $\widetilde{\sigma}_{11}, \ldots ,
\widetilde{\pi}_2$ are too long to be written out in full. The structure
equations for (\ref{HS}) differ from (\ref{SE}) only in replacing $\theta_0,
\ldots , \pi_2$ by their tilded counterparts. Therefore, results of Cartan's
method (see, e.g., \cite[Theorem~15.12]{Olver95}) yield the contact equivalence of
equations (\ref{HS}) and (\ref{EP}). Since the Maurer--Cartan forms for both
symmetry groups are known, the equivalence transformation
$\Psi : (t,x,u,u_t,u_x) \mapsto
(\widetilde{t},\widetilde{x},\widetilde{u},\widetilde{u}_{\widetilde{t}},
\widetilde{u}_{\widetilde{x}})$
can be found from the requirements $\Psi^{*}\widetilde{\theta}_0 = \theta_0$,
$\Psi^{*}\widetilde{\theta}_1 = \theta_1$,
$\Psi^{*}\widetilde{\theta}_2 = \theta_2$,
$\Psi^{*}\widetilde{\xi}^1 = \xi^1$, and $\Psi^{*}\widetilde{\xi}^2 = \xi^2$:

\begin{theorem}
\label{HS_transformation}
The contact transformation $\Psi$
\begin{gather*}
\widetilde{u} =(t+x)^{-\frac{1}{\kappa}}\left(\kappa\,(t+x)\,u_x
+(\kappa-1)\,u\right),
\\
\widetilde{t} = \kappa^{-1}\,t,
\\
\widetilde{x} = -(t+x)^{\frac{\kappa-1}{\kappa}}
\left(\kappa\,(t+x)\,u_x- u\right),
\\
\widetilde{u}_{\widetilde{t}}
= \kappa^2\,(t+x)^{-\frac{1}{\kappa}}\left(u_t-u_x \right),
\\
\widetilde{u}_{\widetilde{x}} = - (t+x)^{-1}
\end{gather*}
takes the Euler--Poisson equation \eqref{EP} to the generalized Hunter--Saxton
equa\-ti\-on \eqref{HS} (writ\-ten in the tilded variables).
\end{theorem}

Equation (\ref{EP}) has an intermediate integral, and its general solution can
be found in quadratures. Indeed, for the invariants of this equation we have
$P + Q = 2$, therefore the Laplace $t$-transformation,
\cite[Section~9.3]{Ovsiannikov}, takes equation (\ref{EP}) to a factorizable linear
hyperbolic equation. Namely, we consider the system
\begin{gather}
v = u_x - (\kappa\,(t+x))^{-1} u,
\label{B1}
\\
v_t = 2\,(1-\kappa)\,(\kappa\,(t+x))^{-1} v + \kappa^{-1}\,(t+x)^{-2} u.
\label{B2}
\end{gather}
Substituting (\ref{B1}) into (\ref{B2}) yields equation (\ref{EP}), while
expressing $u$ from (\ref{B2}) and substituting it into (\ref{B1}) gives the
equation
\begin{gather}
v_{tx} =
  \frac{1- 2\,\kappa}{\kappa\,(t+x)} \, v_t
+ \frac{2\,(\kappa-1)}{\kappa\,(t+x)} \, v_x
- \frac{(2\,\kappa-1)\,(\kappa-2)}{(\kappa\,(t+x))^{2}} v
\label{EP_trans}
\end{gather}
with the trivial Laplace semi-invariant $H$. Hence, the substitution
\begin{gather}
w = v_x + (2\,\kappa-1)\,(\kappa\,(t+x))^{-1} v
\label{B3}
\end{gather}
takes equation (\ref{EP_trans}) into the equation
\begin{gather}
w_t = - 2\,(\kappa-1)\,(\kappa\,(t+x))^{-1} w.
\label{B4}
\end{gather}
Integrating (\ref{B4}) and (\ref{B3}), we have the general solution for
equation (\ref{EP_trans}):
\[
v = (t+x)^{\frac{1-2\kappa}{\kappa}}\,\left(
S(t) + \int R(x)\,(t+x)^{\frac{1}{\kappa}}\,d x
\right),
\]
where $S(t)$ and $R(x)$ are arbitrary smooth functions of their arguments.
Then equation (\ref{B2}) gives the general solution for equation (\ref{EP}):
\[
u{=}
(t{+}x)^{\frac{1}{\kappa}}\,\left(
\kappa\,S^{\prime}(t){+}\int R(x)\,(t{+}x)^{\frac{1{-}\kappa}{\kappa}}\,d x
\right){-}(t{+}x)^{\frac{1{-}\kappa}{\kappa}}\,
\left(
S(t){+}\int R(x)\,(t{+}x)^{\frac{1}{\kappa}}\,d x
\right).
\]
This formula together with the contact transformation from Theorem~\ref{HS_transformation}
gives the general solution for the generalized
Hunter--Saxton equation (\ref{HS}) in a parametric form:
\begin{gather*}
\widetilde{u} = \kappa^2\,S^{\prime}(t)
+\kappa\,\int R(x)\,(t+x)^{\frac{1-\kappa}{\kappa}}\,d x,
\\
\widetilde{t} = \kappa^{-1}\,t,
\\
\widetilde{x} = -\kappa\,
\left( S(t)
+\int R(x)\,(t+x)^{\frac{1}{\kappa}}\,d x\right).
\end{gather*}
Hence,  we obtain the general solution of equation (\ref{HS}) without employing
nonlocal transformations.

\section{Discussion}

Cartan's method of equivalence in its different incarnations is a powerful tool
in the study of symmetry groups of differential equations.  Each of the four
approaches discussed has its own advantages and shortcomings. While the first
and the second approaches are not universal, the third and the fourth
approaches provide an effective technique which is valid irrespective of
geometric properties of a given differential equation or of intransitivity of
the symmetry pseudo-group. The last approach, the moving coframe method, is
applicable to solving the equivalence problems, finding differential invariants
and operators of invariant differentiation for classes of dif\-fe\-ren\-ti\-al
equations where the power of the infinitesimal method is restricted by
complexity of analysis of the defining systems for symmetry generators. Unlike
the second and the third methods, the moving coframe method does not require
analysis of the defining equations and en\-ab\-les one to find the
Maurer--Cartan forms for symmetry pseudo-groups explicitly. The price of its
power is lengthy and massive computations. It would be worthwhile to improve
the al\-go\-rithms of the basic steps involved in the method.

We conclude the paper by mentioning evident prospects for the use of Cartan's
method in studying structure of symmetry pseudo-groups of differential
equations. The directions of fu\-tu\-re research would include, among others,
simplification of solving symmetry classification pro\-blems and clarification of
the group foliation technique with applications to construction of
dif\-fe\-ren\-ti\-al\-ly-invariant solutions. Therefore, the further development and
improvement of im\-ple\-men\-ta\-ti\-ons of Cartan's method would be of
great interest and significance for the study of geometry of differential
equations.
\LastPageEnding
\end{document}